\documentclass[letter]{aa} 

\usepackage{natbib}
\usepackage{indentfirst}
\usepackage[applemac]{inputenc}
\bibpunct{(}{)}{,}{a}{}{,} 
\usepackage{graphicx}
\usepackage{longtable}
\usepackage{txfonts}
\usepackage{color}
\usepackage{soul}
\usepackage{multirow}
\usepackage{textcomp}
\usepackage{ulem}

\newcommand{\kms}{km\,s$^{-1}$}

\newcommand\mic{$\mu$m}

\newcommand{\tcr}[1]{\textcolor{black}{#1}} 

\newcommand{\Figureone}{%
	\begin{figure}[!h]
		\centering
		\includegraphics[scale=0.7]{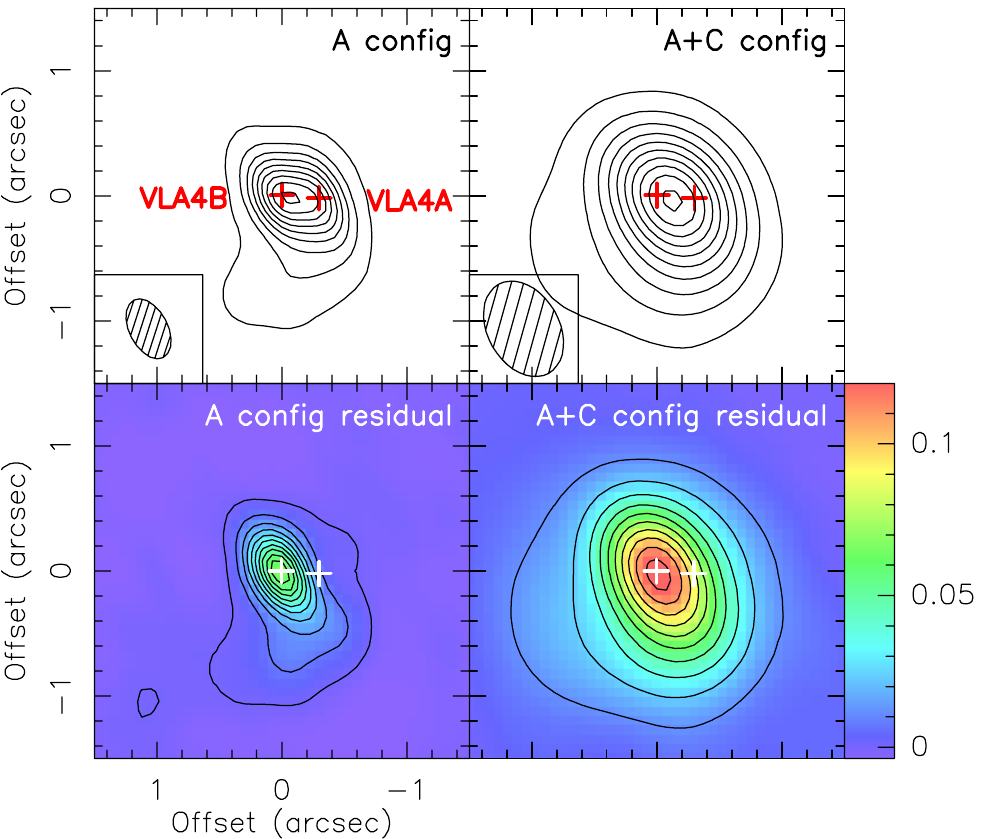}
		\caption{\textit{Top:}  {1.4\,mm} continuum maps of SVS13\,A using  A configuration (left) and A+C (right).  {Contours start at 3$\sigma$ and 10$\sigma$ respectively, and increase by 10$\sigma$.} Crosses show the positions of VLA4B and VLA4A, corrected for our proper motions (Appendix~\ref{app:pmotion}). The synthesized beam is plotted in the \tcr{bottom} left corner. \textit{Bottom}:  {residuals after removal of a point source fitted to the A config. \textit{uv}-data at the position of VLA4A}. 
			The color scale is in mJy/beam.}
		\label{fig:cont1mm}
		\vspace{-0.3cm}
	\end{figure}
}

\newcommand{\Figuretwo}{%
	\begin{figure}[!t]
		\centering
		\includegraphics[scale=0.32]{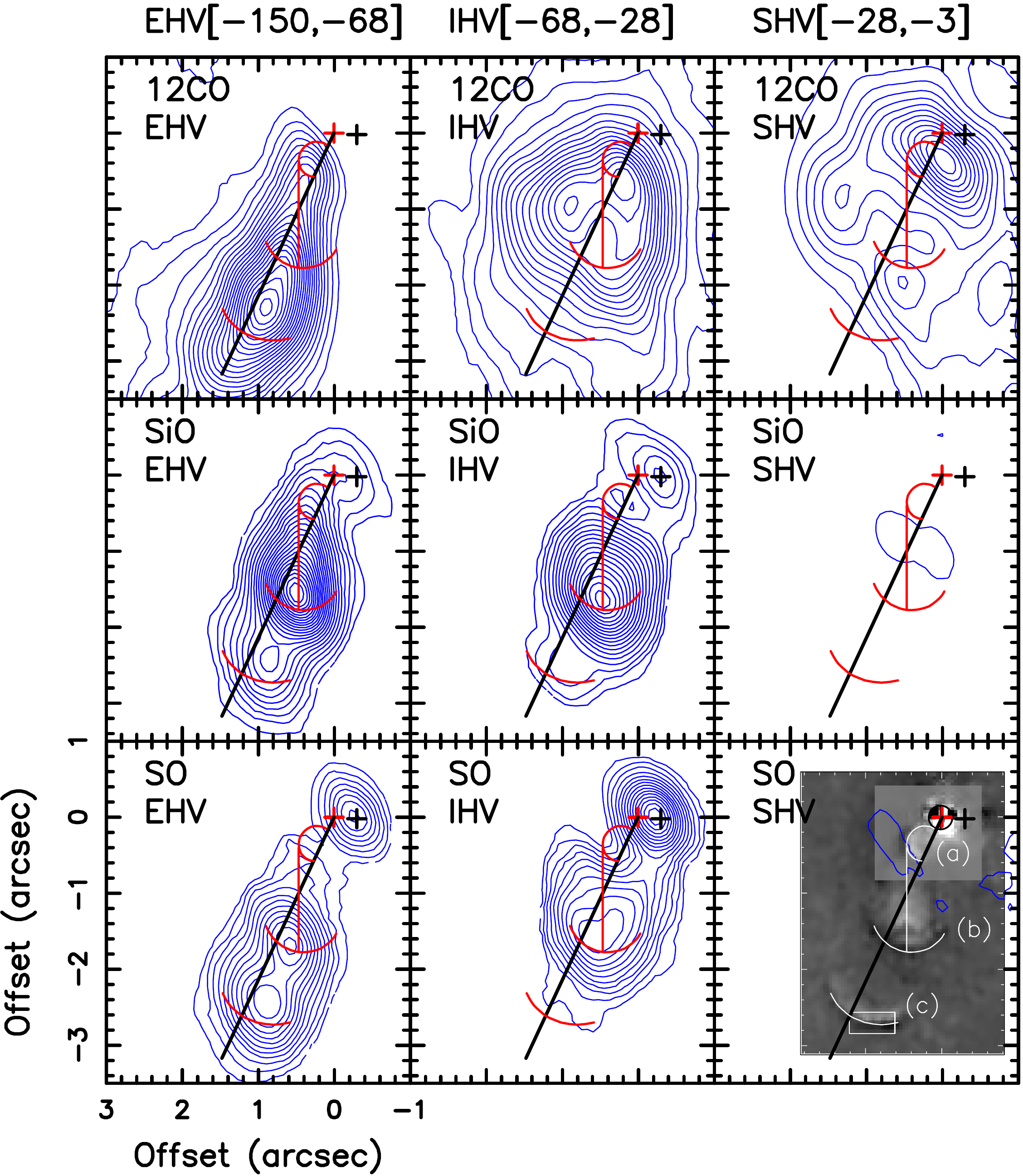}
		\caption{Maps of CO, SiO and SO integrated over three $V_{\rm LSR}$  ranges: EHV\,=\,-150 to -68\,km\,s$^{-1}$, IHV\,=\,-68 to -28\,km\,s$^{-1}$, SHV\,=\,-28 to -3\,km\,s$^{-1}$. Red and black crosses denote VLA4B and VLA4A, respectively. 
			\tcr{The compact emission peaked on VLA4A is entirely due to COMs (identified in Figs.~\ref{fig:Hodapp_SiO} and \ref{fig:Hodapp_SO}).} 
			The bottom right panel shows Fig.~4 of HC14 registered on VLA4B.  Their main H$_2$ features are outlined in white \tcr{(cf. text and expanded view in Fig.~\ref{fig:Hodapp_zoom}).} 
			They are reported in red in the other panels. The black line displays PA=155\degr (B00). }
		\label{fig:EHV-IHV-SHV}
		\vspace{-0.3cm}
	\end{figure}
}

\newcommand{\Figurethree}{%
	\begin{figure}[t!]
		\centering
		\includegraphics[scale=0.5]{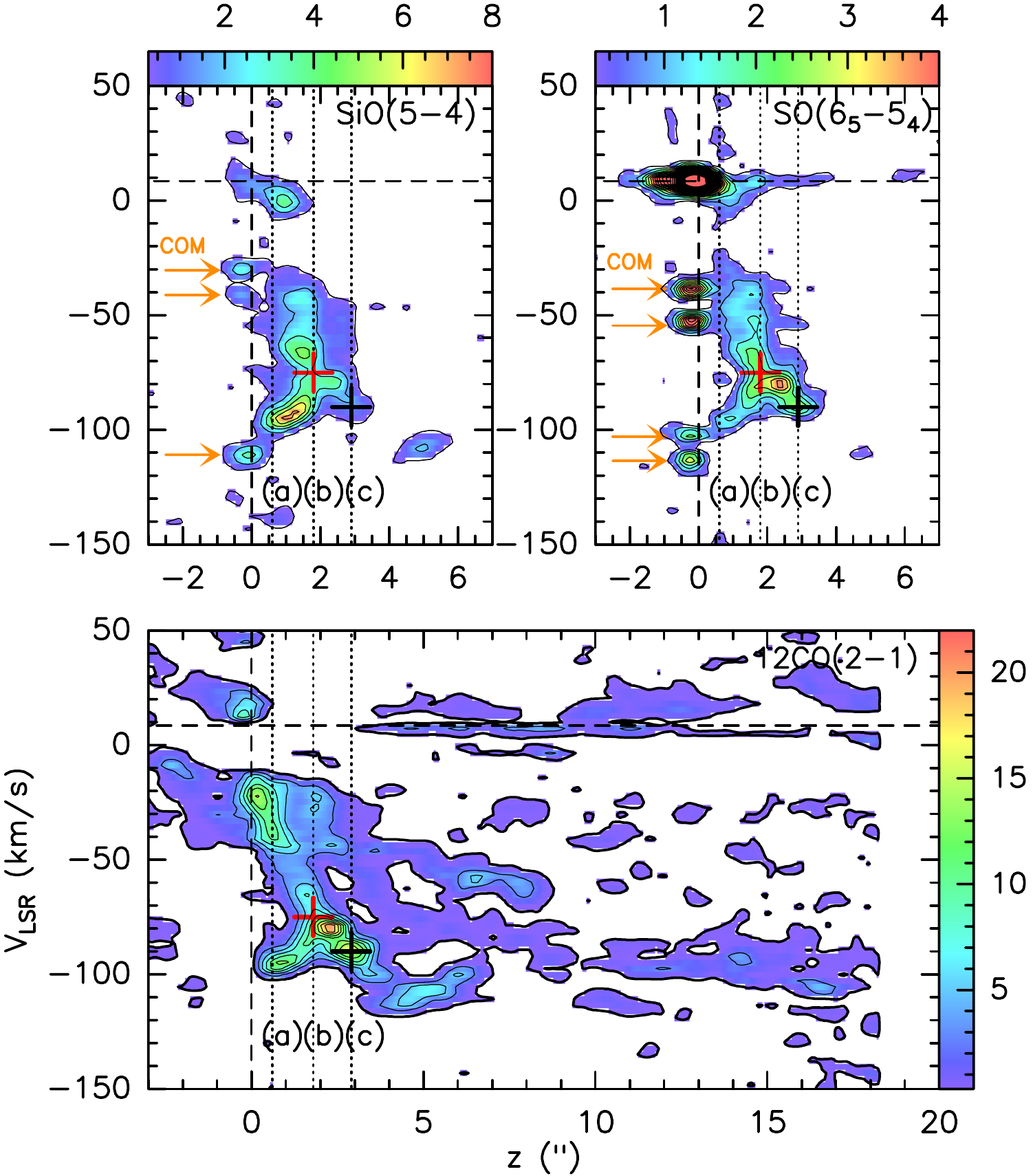}
		\caption{PV diagrams along PA=155\textdegree. Dashed lines indicate the reference position (VLA\,4B) and the systemic velocity  {V$_\mathrm{sys}$\,=\,8.5\,\kms} \tcr{(from SO)}. Contours start at 3$\sigma$ {with steps of} 14$\sigma$ for SiO(5--4),  7$\sigma$ for SO($6_5-5_4$) and 21$\sigma$ for CO(2--1). The color scale displays the intensity in K.  Vertical dotted lines mark the position of the H$_2$ features of HC14 as labelled in Fig.\,\ref{fig:EHV-IHV-SHV}. {Red and black crosses show velocities for which the molecules peak on arcs \textit{(b)} and \textit{(c)}, respectively. The orange arrows indicate contamination by \tcr{complex organic} molecules \tcr{(COMs)}.}}
		\label{fig:pv}
				\vspace{-0.3cm}
	\end{figure}
}

\newcommand{\Figurefour}{%
	\begin{figure}[t!]
		\centering
		\includegraphics[scale=0.4]{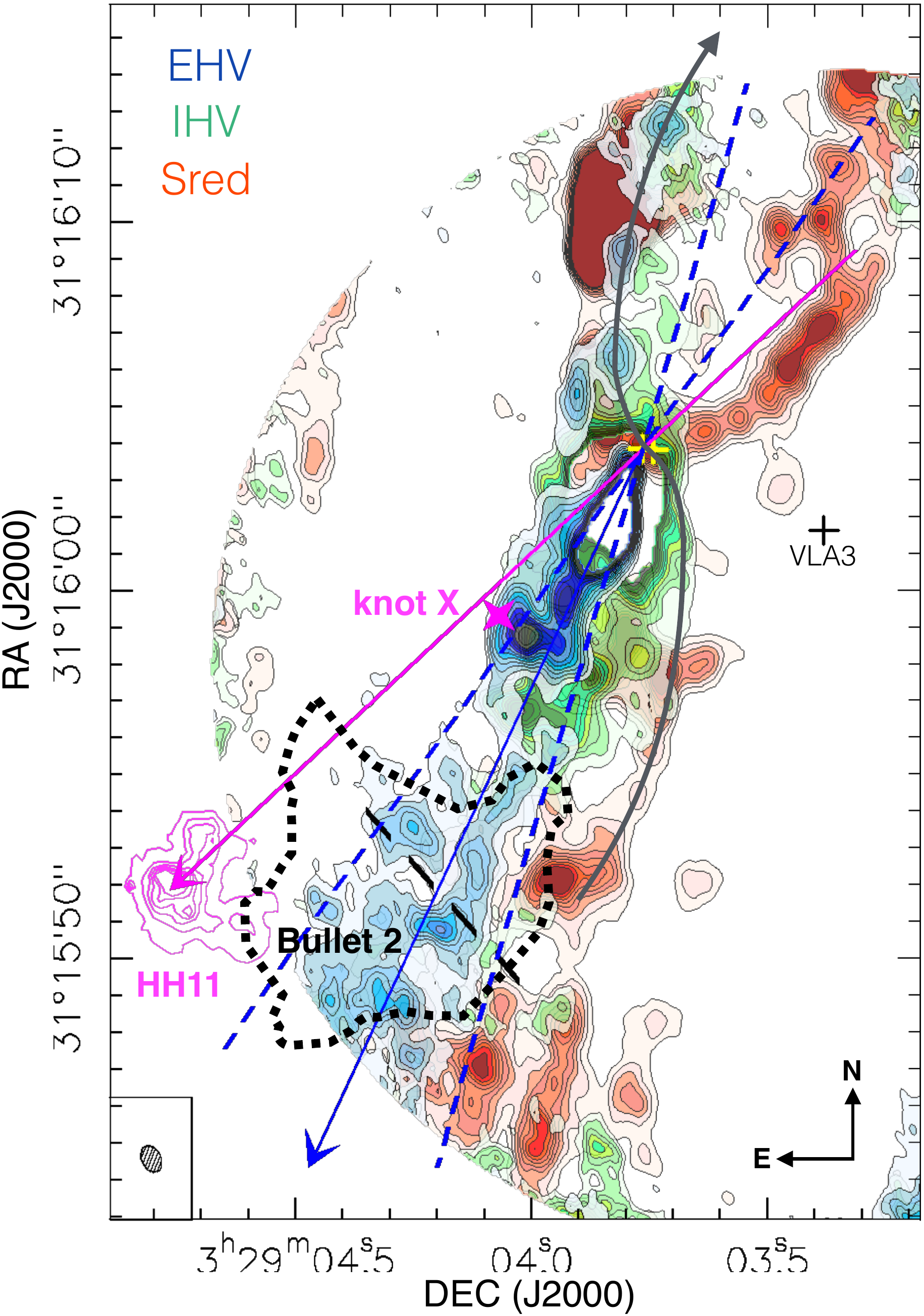}
		\caption{Wide field CO maps of the SVS\,13A outflow  {corrected from primary beam attenuation}, in three velocity ranges: blue:\,EHV\,(-150 to -68\,\tcr{\kms}), green:\,IHV\,(-68 to -28\,\tcr{\kms}), red:\,Sred\,(+10 to +40\,\tcr{\kms}). The three arrows {show} \tcr{the directions of the three jets: to HH11 (PA=\tcr{133}\degr, pink), the {wiggling} H$_2$/CO jet (PA=155\,$\pm$\,10\degr, solid and dashed blue lines) and the third jet suggested by {velocity} sign reversal (PA$\simeq$0\degr, grey).}	
			The \tcr{pink} cross marks the SiO/SO knot X at -28\,\kms. \tcr{Pink contours show HH11 extension \citep{2001Noriega}.}
			Dotted black contours outline the EHV Bullet\,2 as mapped by {B00}. Dashed segment shows where \citet{2016Chen} reported possible rotation. VLA4A/4B are marked as yellow crosses.}
		\label{fig:12CO_widefield}
		\vspace{-0.5cm}
	\end{figure}
}

\newcommand{\FigSiOcentroid}{%
	\begin{figure}[t!]
		\centering
		\includegraphics[scale=0.6]{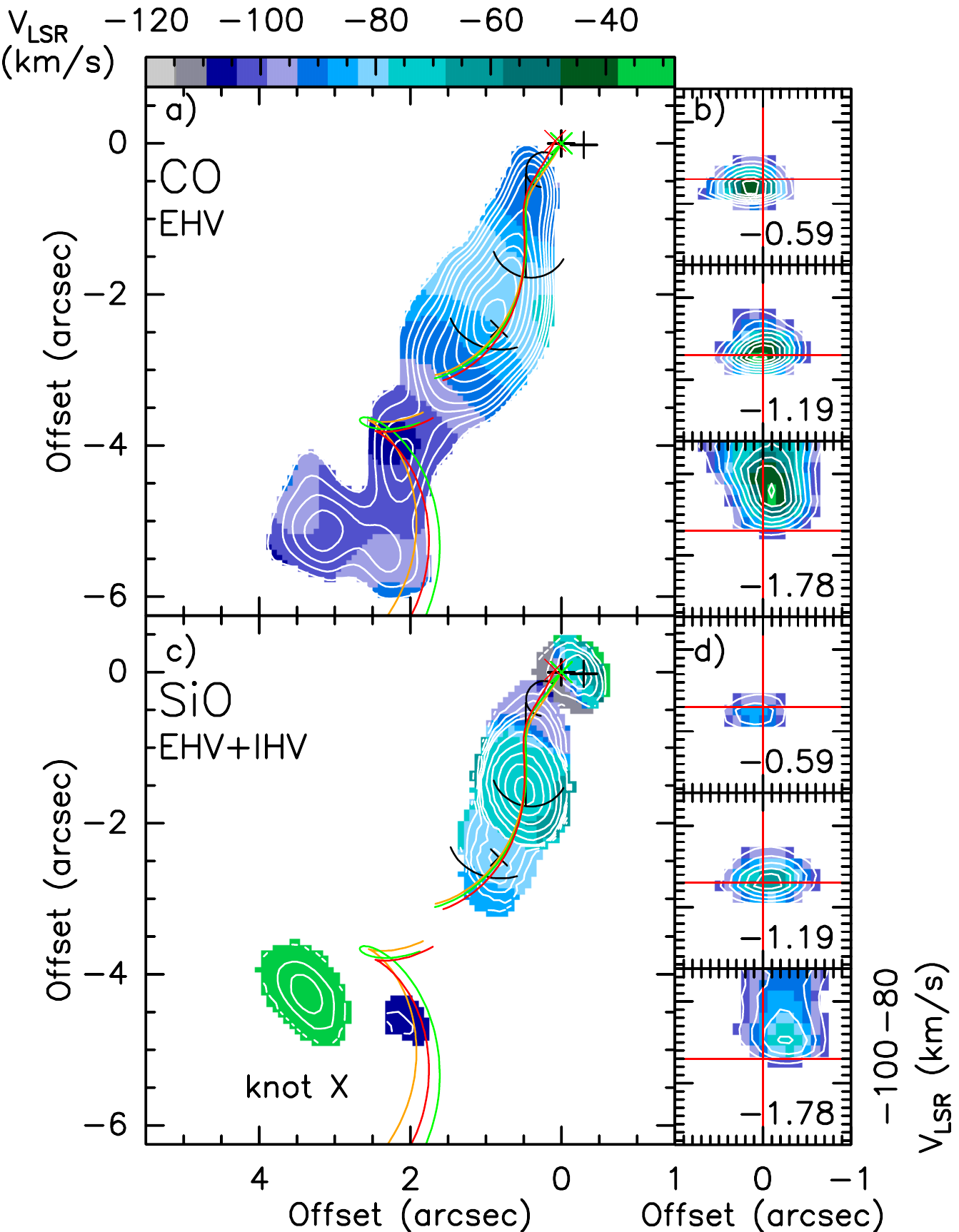}
		\caption{Left panels: \tcr{Centroid velocity (colour) and integrated intensity (contours)} in CO(2--1) \tcr{EHV}  (a) and SiO(5--4) \tcr{EHV+IHV}  (c). \tcr{Best fit} \tcr{wiggling models are overplotted in red, orange, and green (see App. C for details)}. Right panels: transverse PV diagrams of
			CO(2--1) (b) and SiO(5--4) (d) at  {the three distances marked} in each sub-panel, \tcr{and perpendicular to mean PA=155\degr}.  {Red lines are visual cues.} 
		}
		\label{fig:siocentroid}
		\vspace{-0.5cm}
	\end{figure}
}

\newcommand{\Figpmotion}{%
	\begin{figure}[h!]
		\centering
		\vspace{-1.cm}
		\includegraphics[scale=.28]{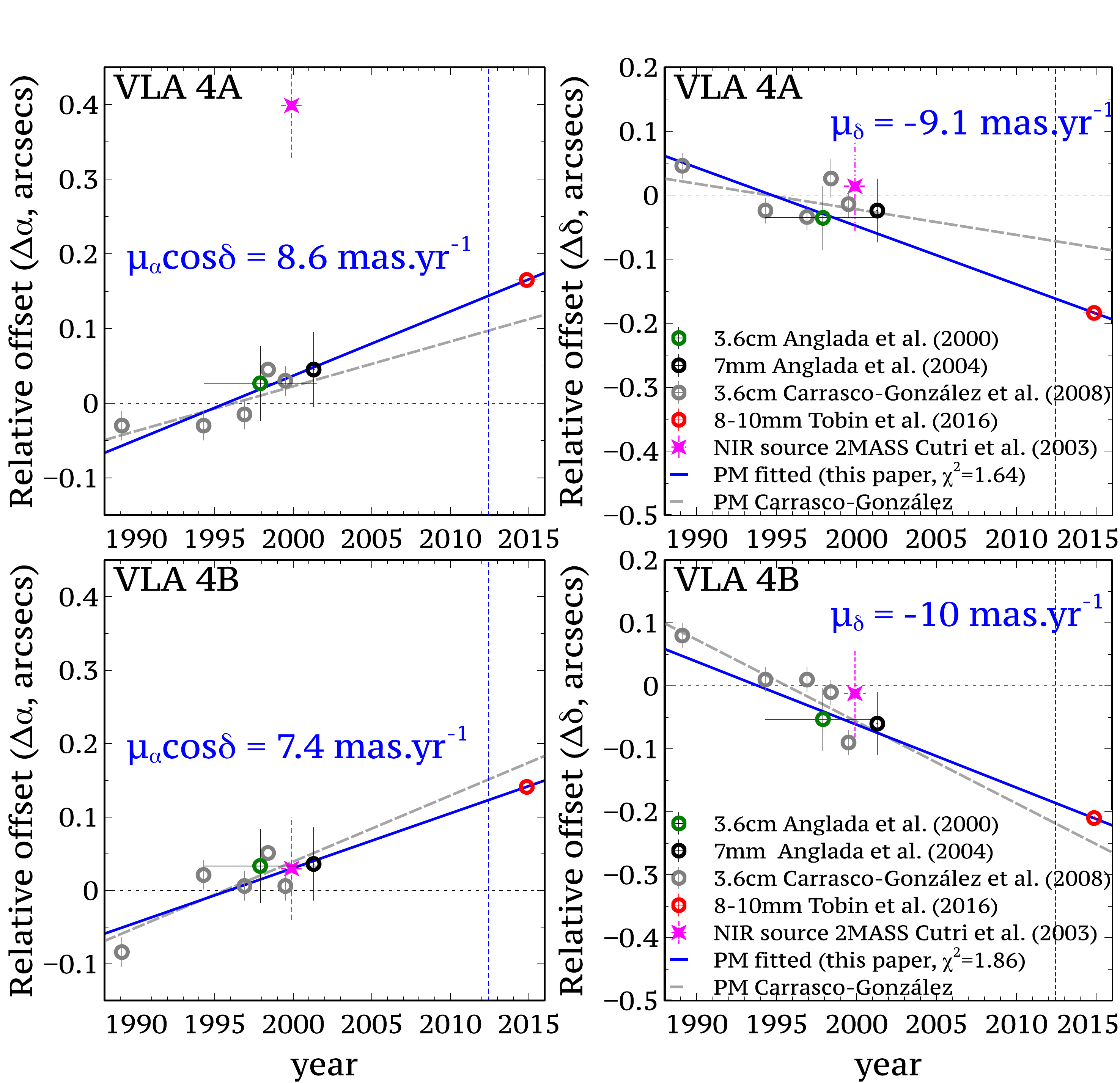}
		\caption{Revised proper motions of VLA4A and VLA4B. 
			This figure is identical to Fig.~2 of CG08 with additional data. Reference positions are 03:29:03.732 31:16:03.974 (J2000) for VLA4A, and 03:29:03.7566 31:16:04.000 (J2000) for VLA4B. Dotted grey lines display CG08 proper motions based on their measurements (grey points).
			Solid blue lines are our best linear fit to all VLA radio data including the positions of \cite{2016Tobin}. Vertical blue dashed lines identify the weighted date of CALYPSO observations (2012.43).}
		\label{fig:pmotion}
		\vspace{-0.3cm}
	\end{figure}
}

\newcommand{\Figuretwoapp}{%
	\begin{figure}[!t]
		\centering
		\includegraphics[scale=0.5]{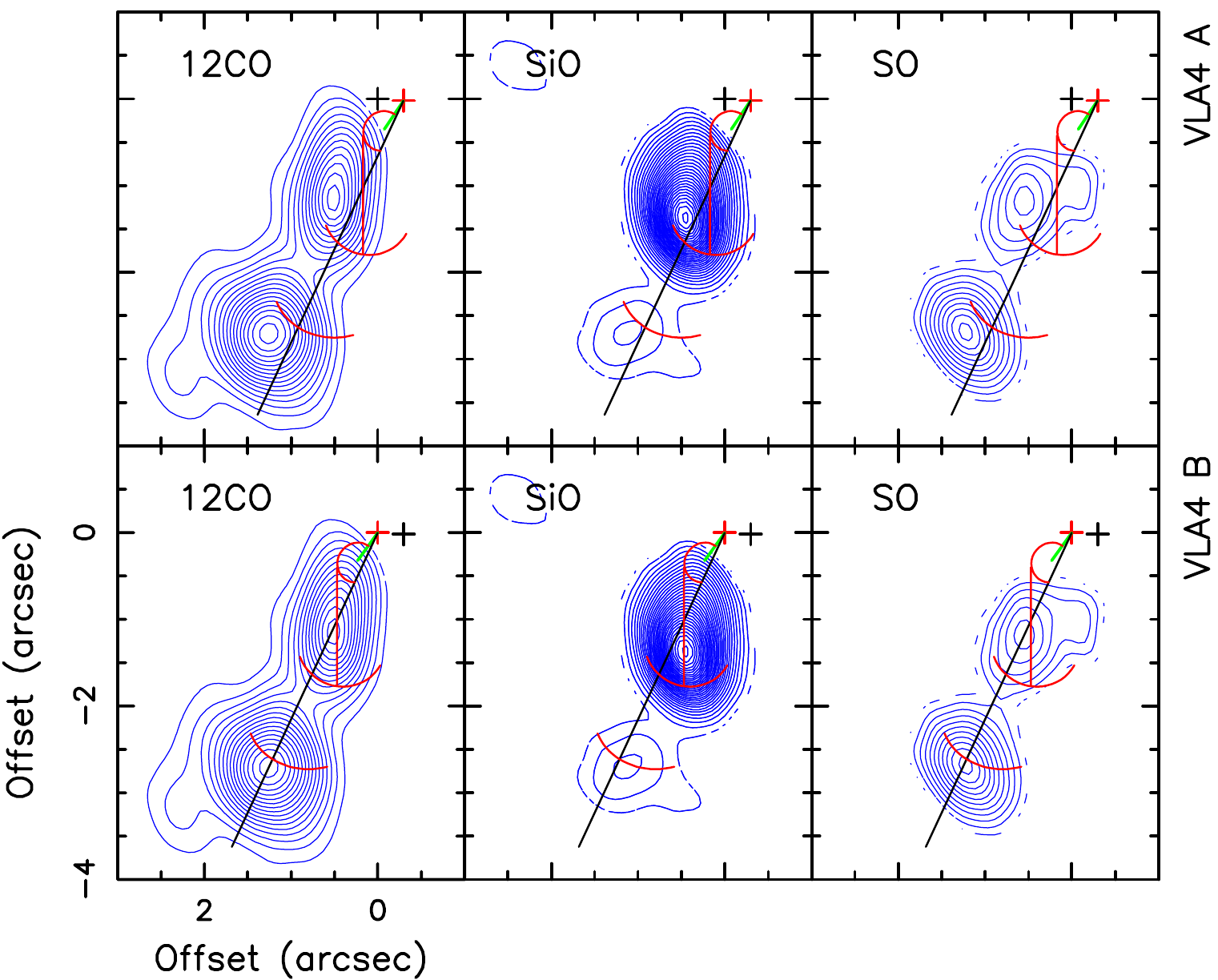}
		\caption{Channel maps of CO, SiO and SO at $V_\mathrm{LSR}=$\,-92.15\,\kms\ (blue contours) with the positions of the [FeII] jet (green segment), H$_2$ inner bubble, arcs and vertical linear feature (in red) from HC14 superimposed. The NIR source was assumed to be either VLA4A (top row) or VLA4B (bottom row). The latter is unambiguously favored.}
		\label{fig:Hodapp_EHV}
		\vspace{-0.3cm}		
	\end{figure}
}

\newcommand{\FiguretwoappHodapp}{%
	\begin{figure}[t!]
		\centering
		\includegraphics[scale=0.35]{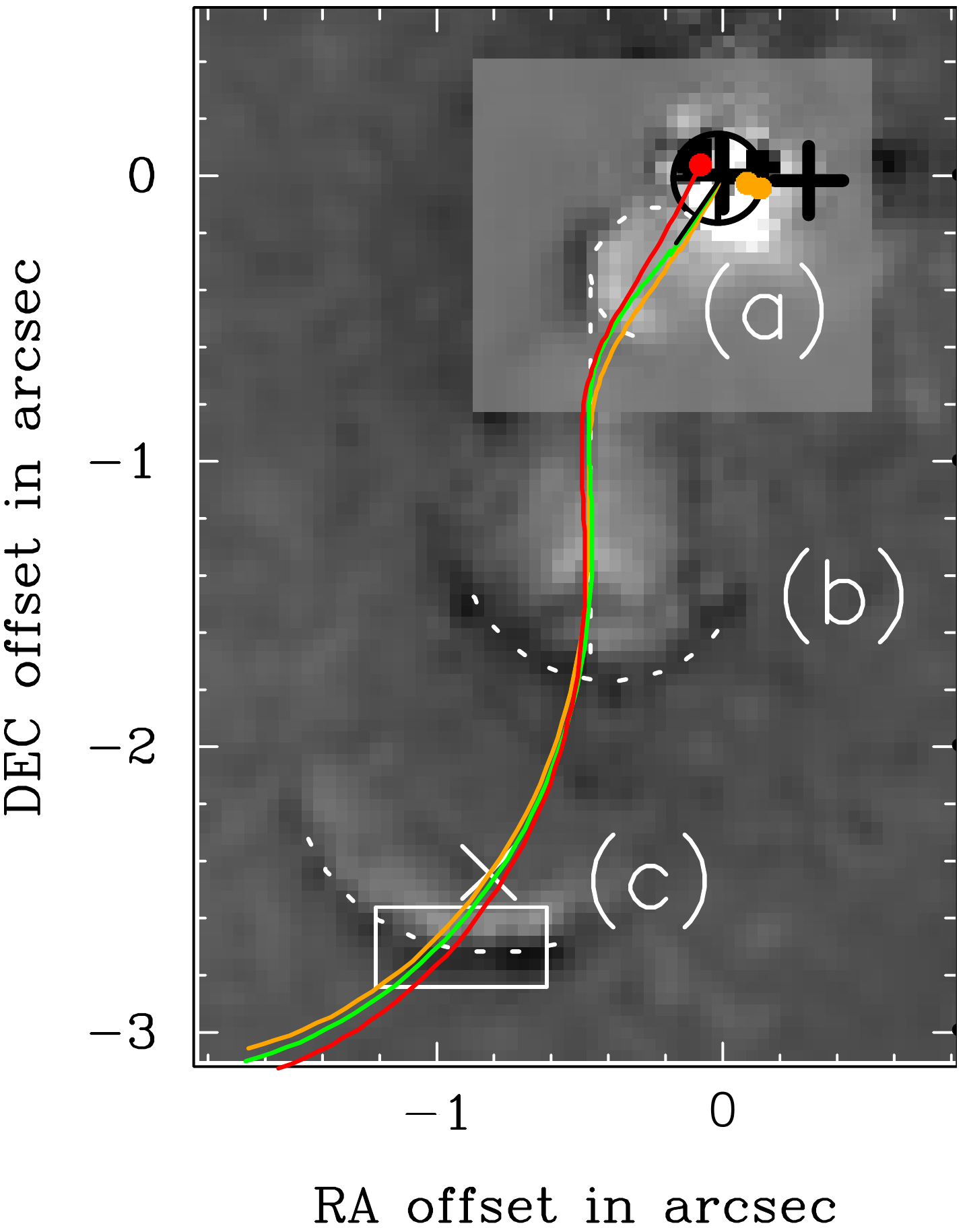}
		\caption{Fig.~4 of HC14 registered on VLA4B, {with their main H$_2$ features outlined in dashed white: inner "bubble"
			(denoted \textit{(a)}), outer arcs (denoted \textit{(b)} and \textit{(c)}), 
			and the fainter "linear feature" joining \textit{(a)} and \textit{(b)}}. {A cross denotes the SiO EHV peak ahead of arc \textit{(c)}}. {Coloured curves show our best wiggling models in three scenarios} (see Table \ref{tab:wiggling} for details). {Predicted companion} positions {in the} orbital {models} are displayed by filled circles of same colour. {The [FeII]} jet from VLA4B (cf. HC14) is shown {as a short black segment}.}
		\label{fig:Hodapp_zoom}
		\vspace{-0.8cm}
	\end{figure}
}

\newcommand{\FigHodappCO}{%
	\begin{figure*}[t!]
		\centering
		\includegraphics[scale=1.0]{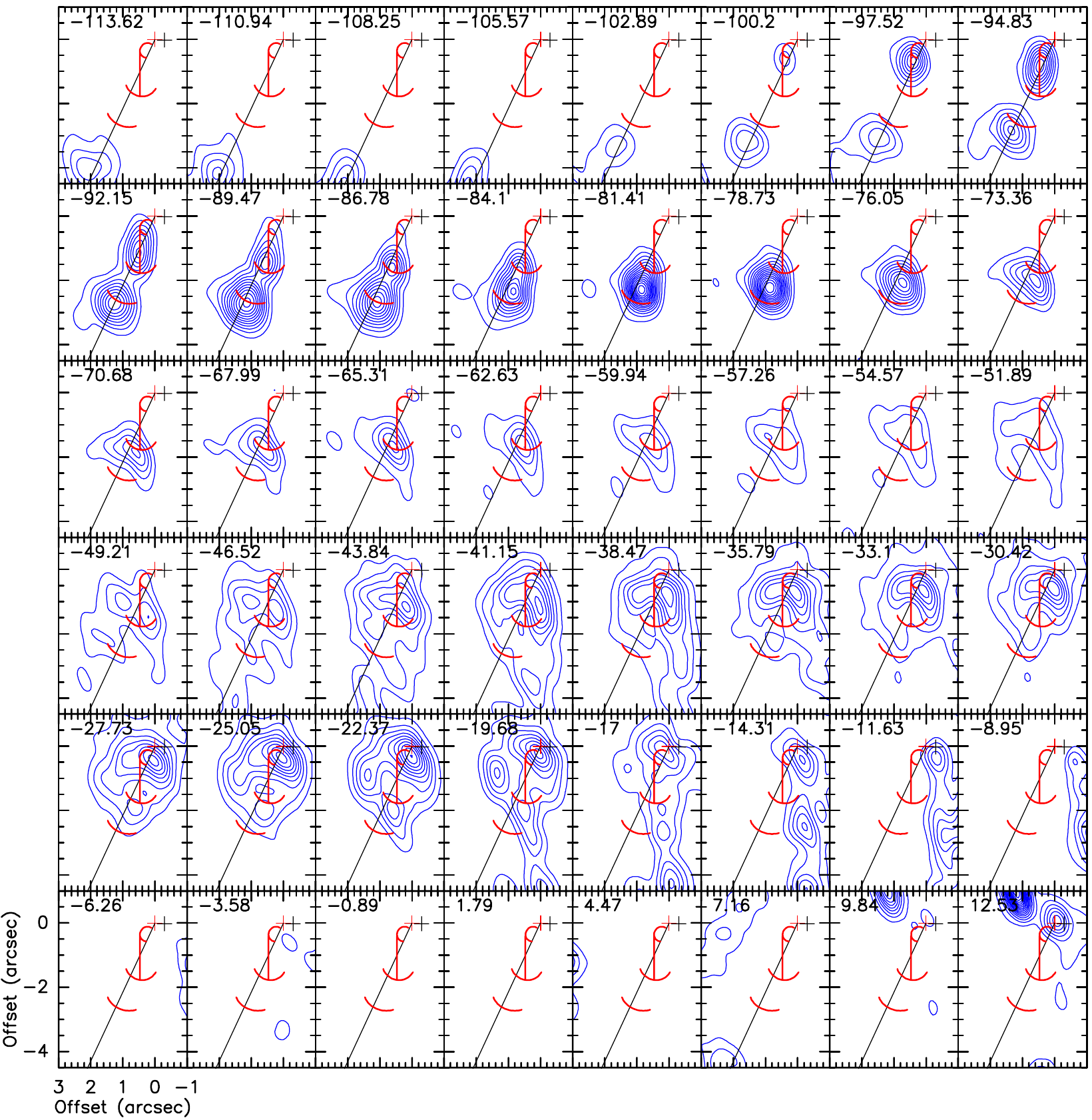}
		\caption{CO(2--1) channel maps over the same field of view as in Fig.~\ref{fig:EHV-IHV-SHV} with H$_2$ features from HC14 drawn in red, and PA=155\degr\ shown as a black line of length 5\arcsec. Red and black crosses mark VLA4B (east) and VLA4A (west), respectively.}
		\label{fig:Hodapp_12CO}
	\end{figure*}
}

\newcommand{\FigHodappSO}{%
	\begin{figure*}[t!]
		\centering
		\includegraphics[scale=1.0]{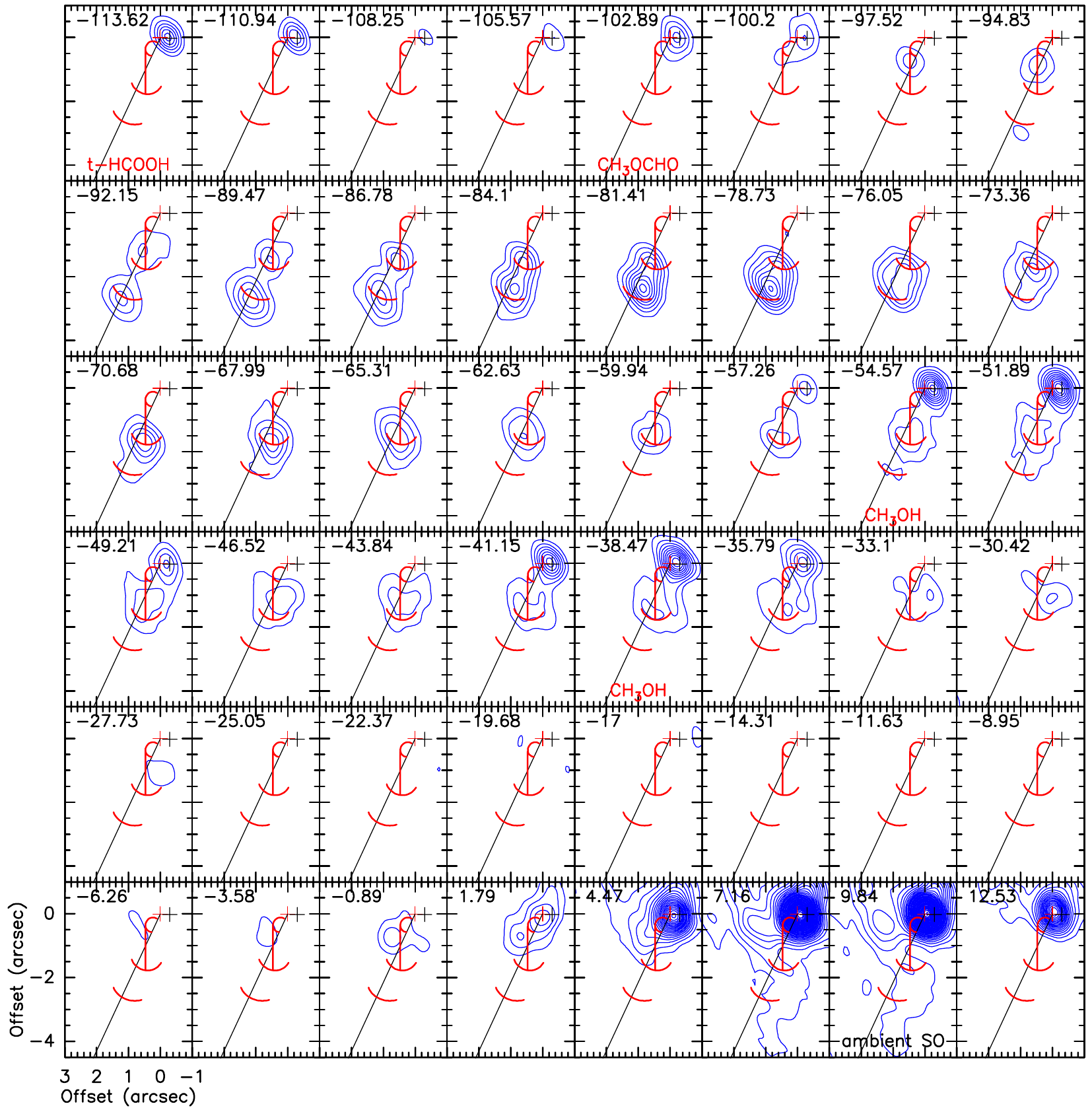}
		\caption{SO($6_5-5_4$) channel maps over the same field of view as in Fig.~\ref{fig:EHV-IHV-SHV} with H$_2$ features from HC14 drawn in red, and PA=155\degr\ shown as a black line of length 5\arcsec. Red and black crosses mark VLA4B (east) and VLA4A (west), respectively. Complex organic molecules present in this range are identified by red labels in their central emission channel (from Belloche et al., in prep).}
		\label{fig:Hodapp_SO}
	\end{figure*}
}

\newcommand{\FigHodappSiO}{%
	\begin{figure*}[t!]
		\centering
		\includegraphics[scale=1.0]{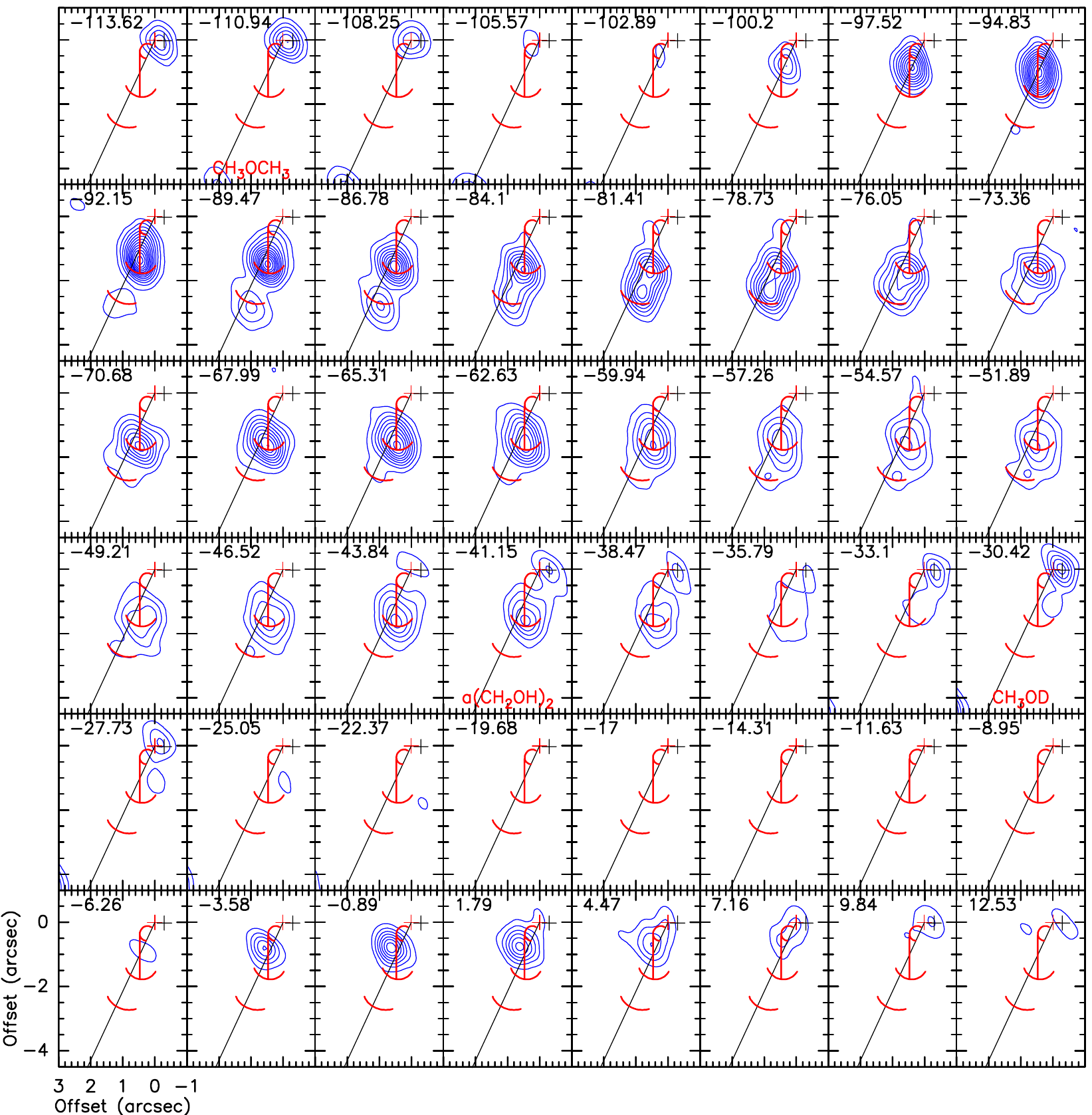}
		\caption{SiO(5--4) channel maps over the same field of view as in Fig.~\ref{fig:EHV-IHV-SHV} with H$_2$ features from HC14 drawn in red, and PA=155\degr\ shown as a black line of length 5\arcsec. Red and black crosses mark VLA4B (east) and VLA4A (west), respectively. Complex organic molecules present in this range are identified by red labels in their central emission channel (from Belloche et al., in prep).}
		\label{fig:Hodapp_SiO}
	\end{figure*}
}

\begin{document}
	
	\title{CALYPSO view of SVS\,13A with PdBI: Multiple jet sources}
	
	\titlerunning{CALYPSO view of SVS13\,A}

	\authorrunning{C. Lef{\`e}vre et al.}		
	\author{C. Lef{\`e}vre  \inst{1}, S. Cabrit  \inst{2},
		A. J. Maury \inst{3}, F. Gueth \inst{1}, B. Tabone  \inst{2}, L. Podio \inst{4}, A. Belloche\inst{5}, C. Codella \inst{4}, S. Maret \inst{6}, S. Anderl \inst{6}, Ph. Andr{\'e} \inst{3}, P. Hennebelle \inst{3}}
	
	\offprints{lefevre@iram.fr}
	
	\institute{Institut de Radioastronomie Millim\'etrique
		(IRAM), 300 rue de la Piscine, 38406 Saint-Martin
		d'H\`eres, France
		\and
		LERMA, Observatoire de Paris, CNRS, 61 Av. de l'Observatoire, 75014 Paris, France				
		\and
		Laboratoire AIM-Paris-Saclay, CEA/DSM/Irfu - CNRS - Universit\'e Paris Diderot, CE-Saclay, 91191     Gif-sur-Yvette, France
		\and
		Osservatorio Astrofisico di Arcetri, Largo Enrico Fermi 5, I-50125 Florence, Italy
		\and		
		Max-Planck-Institut f\"ur Radioastronomie, Auf dem H\"ugel 69, 53121 Bonn, Germany
		\and
		Univ. Grenoble Alpes, CNRS, IPAG, F-38000 Grenoble, France
	}
	
	\date{Received March 10, 2017; accepted July 6, 2017}
	
	\abstract
	{}
	{We wish to clarify the origin of \tcr{the multiple jet features emanating from the binary protostar SVS\,13A (=\,VLA4A/VLA4B).}}
	{We used the Plateau de Bure Interferometer to map at 0.3--0.8\arcsec\ \tcr{($\sim$70--190 au)} dust emission at 1.4\,mm, CO(2--1), SiO(5--4), SO(6$_5$--5$_4$). Revised proper motions for VLA4A/4B \tcr{and jet wiggling models} are computed to clarify their respective contribution.}
	{VLA4A shows compact dust emission suggestive of a disk\,$<$\,50\,au, and is the hot corino source\tcr{, while CO/SiO/SO counterparts to the small-scale H$_2$ jet originate from VLA4B and reveal the jet {variable} velocity structure. This jet exhibits \tcr{$\simeq$\,3$\arcsec$\,} wiggling consistent with orbital motion around a yet undetected $\simeq$\,20--30\,au companion to VLA4B, or jet precession. Jet wiggling combined with velocity variability can explain the large apparent angular momentum in CO bullets.} We {also} uncover a synchronicity between CO jet bullets and knots in the HH7--11 chain demonstrating that they trace two distinct jets. Their $\simeq$\,300\,yr twin outburst period may be triggered by close perihelion approach of VLA4A in an eccentric orbit around VLA4B. A third jet \tcr{is tentatively seen} at PA\,$\simeq$\,0$\degr$.}
	{\tcr{SVS13\,A harbors at least 2 and possibly 3 distinct jet sources. The CO and HH7-11 jets are launched from quasi-coplanar disks, separated by 20--70~au.} Their synchronous major events every 300\,yr favor \tcr{external triggering by close binary interactions, a scenario also invoked for FU~Or outbursts.} }

	
	\maketitle
	

	\section{Introduction}
		\vspace{-0.1cm}
	
	\tcr{The role of close stellar encounters in accretion and ejection bursts in young stars is a debated issue \citep[and references therein]{2014Reipurth,2014Audard}.} 
	\tcr{An interesting target of study in this context is \object{SVS\,13A}, a 0.3$\arcsec$\,(70\,au) binary} solar-type protostar \citep[\object{VLA4A},\object{VLA4B};][]{2000Anglada} in \object{NGC\,1333} \citep[$d$\,$\simeq$\,235\,pc,][]{2008Hirota} with \tcr{multiple signs of ejection bursts.} SVS\,13A is the candidate source of the \object{HH7--11} chain of optical\tcr{/H$_2$ knots \citep{2000Davis}}, drives {high-velocity} CO bullets \citep[][hereafter B00]{2000Bachiller}, and hosts small scale [Fe~II] jet with H$_2$ arcs \citep[see][hereafter HC14, and references therein]{2014Hodapp}. 
	\tcr{An unusually large angular momentum extraction by CO outbursts, compared to other rotating outflows like DG Tau,} was recently claimed from SMA maps at 3\arcsec\ {resolution} \citep{2016Chen}.
\tcr{Here} we present {the highest angular resolution}  (0.3\arcsec-- 0.8\arcsec) study of the SVS13\,A system in dust continuum, CO, SiO and SO, obtained in the frame of the CALYPSO (\textit{Continuum And Lines in Young ProtoStellar Objects}\footnote{\tcr{http://irfu.cea.fr/Projets/Calypso}}) Large Program at the \tcr{IRAM} Plateau de Bure \tcr{inteferometer} (hereafter PdBI, Maury et al. in prep). The high astrometric precision, {angular} resolution, and  {sensitivity} reveal the {jet} structure in unprecedented details and \tcr{the key role of multiplicity in its outstanding outflow properties}.
			
		\vspace{-0.7cm}		
		\section{Observations and data reduction}
		\vspace{-0.2cm}		
		\indent CALYPSO data {of SVS\,13A} were taken with the 6 antennas 
		
		\noindent of {the PdBI} in A and C configurations in the winters of 2011, 2012, and \tcr{2013,} with baselines ranging from 24\,m to 760\,m. 
		The typical system temperature  {and water vapor were 140 K and $\sim$\,2\,mm. } 
		We used the WideX backend 	
		with 2\,MHz channels ({$\sim$2.8\,km\,s$^{-1}$}) to extract a range of $\pm$\,300\kms\ around SiO(5--4), SO($6_5-5_4$) and CO(2--1) at 217.10498, 219.949442, and
		230.53800 GHz\tcr{, respectively}. Data calibration was performed with the GILDAS--CLIC software. 
		Continuum was obtained by averaging line--free channels over the 4\,GHz WideX band centered  {at 219 GHz}.  
		{Self-calibration was performed on the continuum $uv$-data.} 
		The same gain tables were then applied to all WideX $uv$-tables, and {images} cleaned using
		{\tcr{natural weighting}}, 
		with a synthesized beam of 0.83$\arcsec \times$ 0.56$\arcsec$ {in continuum, SiO and SO}
		and 0.76$\arcsec \times$ 0.5$\arcsec$ in CO. 
		Each line cube was continuum-subtracted by baseline fitting
		in the image plane, and also corrected from primary beam attenuation. 
		A high-resolution
		continuum image 
		was generated from A configuration data alone \tcr{with uniform weighting}, resulting  {in a} 0.52$\arcsec \times$ 0.31$\arcsec$ beam. The absolute astrometric precision based on phase calibrators is $\simeq$\,20\,mas. Absolute flux calibration was based on MWC349, with an uncertainty of 20\%.  {The 1$\sigma$ noise level is 0.7 and 1.2\,mJy/beam in continuum for A and A+C configurations, respectively, and 1.7\,mJy/beam\,$\simeq$\,0.1\,K \tcr{per channel} for the lines.} 
		
		\vspace{-0.2cm}	
		\section{Results}
		
		\subsection{ \tcr{Dust and COM distribution in SVS\,13A}}
		\label{sec:dust}
		\Figureone	
		
		\noindent The {1.4\,mm} continuum maps are presented in Fig.\,\ref{fig:cont1mm}. 
		The A-configuration map reveals an EW elongation indicating 
		\tcr{similar peak fluxes towards both sources of the close (0.3\arcsec) binary system VLA4A/B}, marginally resolved by our longest baselines. To estimate the flux of each component,
		we computed updated proper motions {including the latest data} of \cite{2016Tobin} and deduced accurate positions at the time of the CALYPSO observations (Appendix \ref{app:pmotion}). 
		By fitting \tcr{source models} at these positions to the \tcr{A and A+C} \textit{uv}-data, we find that \tcr{VLA4B is extended, while VLA4A can be reproduced by an unresolved component alone (radius $<$\,0.2\arcsec\,$\simeq$\,50\,au)} of flux $\simeq$\,40\,$\pm$\,10\,mJy (where error bars include position and calibration uncertainties).\tcr{The free-free contamination of VLA4A at 1.4\,mm, extrapolating the 3.6--1\,cm fluxes of \cite{2000Anglada} and \cite{2016Tobin}, is only $\simeq$\,1.5\,mJy}. Hence, we find that VLA4A harbors {compact circumstellar dust, suggestive of a disk of radius $<$\,50\,au, unlike previously believed \citep{2004Anglada}.} {Assuming optically thin emission from compact disk or inner envelope, and using Eqs.~(1) and (2) of \citet{2001Motte} with \tcr{L$_\star$\,=\,28\,L$_\odot$ \citep[from Herschel 70\,\mic,][]{2010Andre},}} we estimate an inner circumstellar mass of $\simeq$\,0.01\,$M_\odot$ (T$_\mathrm{dust}$\,$\sim$\,40-100\,K at 50\,au, with uncertainties of a factor three mostly from dust emissivity). \tcr{Interestingly, complex organic molecules (COMs) from the hot corino of SVS13A {\citep[eg.][]{2015LopezSepulcre,2017DeSimone}}
		peak on or close to VLA4A (see Fig.~\ref{fig:EHV-IHV-SHV} and Belloche et al., in prep.), confirming VLA4A has dense circumstellar material.} Residuals after removing the point source on VLA4A (Fig.~\ref{fig:cont1mm}, \tcr{bottom row}) show \tcr{dust} emission extended on scales \tcr{of at least $\sim$2$\arcsec$} peaking on VLA4B (total flux $\sim$\,200\,$\pm$\,25\,mJy, \tcr{free-free contamination $<$\,10\,mJy}). This component may be reproduced by the inner parts of a larger circumbinary envelope (with details to be developed in Maury et al., in prep). 	
			\vspace{-0.15cm}
		
		\subsection{ {Origin \tcr{and kinematics} of the \tcr{blueshifted} molecular jet}}\label{sect:ori_kine}

		\Figuretwo

		{Fig.~\ref{fig:EHV-IHV-SHV}\tcr{, left column,} shows that the extremely-high velocity (hereafter EHV) \tcr{blueshifted (with respect to V$_\mathrm{sys}$\,=\,8.5\,\kms)} CO jet found by B00 
		is resolved into a series of knots, also detected in SiO and SO. {It also reveals} an excellent correspondence with the H$_2$ outflow features of HC14 \tcr{-- namely a small circular "bubble"
			(denoted \textit{(a)}), two H$_2$ arcs (denoted \textit{(b)} and \textit{(c)}), 
			and a fainter "linear feature" joining \textit{(a)} and \textit{(b)} --} when these features are registered on {VLA4B {(see Fig.~\ref{fig:EHV-IHV-SHV})}. 
	In contrast, a systematic shift appears if they are registered on VLA4A (see Appendix \ref{app:jetorigin}). 
	Hence, the CALYPSO data unambiguously 
	identify VLA4B as the \tcr{base} of the small-scale CO\,/\,H$_2$ jet from SVS\,13A,  and hence
	confirming \tcr{previous} assumption\footnote{based on non-simultaneous 2MASS astrometry  {from 1999}}  by HC14
	that VLA4B was the \tcr{near-infrared (NIR)}  source in their observations. 
	The spatial \tcr{agreement} between {tracers with}
	very different excitation conditions (2000\,K for H$_2$, less than a few 100\,K for CALYPSO) implies emission in shock fronts with strong temperature gradients below our {angular} resolution. 
	CALYPSO data {thus reveal} the kinematics of H$_2$ features at unprecedented spectral resolution \tcr{(apart from bubble (\textit{a}) that
		is strongly beam diluted)}.
	
	Position-Velocity (PV) diagrams  {along the mean jet axis (PA=155\textdegree, B00)} are shown in Fig.~\ref{fig:pv}, 
	and individual channel maps in Appendix~\ref{app:cmaps}.
	The EHV jet velocity fluctuates strongly with distance. Along the linear H$_2$ feature
	it shows apparent deceleration from $V_{LSR} \simeq$ -100 \kms\ near bubble (\textit{a}) to  -75$\,\pm\,$5 \kms\ at arc (\textit{b}), 
	then increases again to -90 $\pm$\,3\,\kms\ at arc (\textit{c}).
	{Combined with the arc (\textit{c}) proper motion of 31.4\,mas/yr\,=\,35\,\kms\ (HC14) we infer an inclination $i$\,$\simeq$\,{20}$\degr$ for this feature, similar to the center of the inner bubble ($i$\,$\simeq$\,18$\degr$, HC14).}
	{The jet velocity further increases beyond arc (\textit{c}) \tcr{up to -115 \kms\ at 4\arcsec\, and oscillates between {-115} and -90\,\kms\ on larger scale (see Fig.~\ref{fig:pv}). No redshifted EHV counterjet is detected, as noted by B00.}
		\Figurethree
		
		{The intermediate and standard high--velocity ranges (IHV, SHV) 
			show an apparent accelerating pattern from \tcr{$\simeq$\,-20\,\kms}  near the source to -70\,\kms\ at arc (\textit{b}) (Fig.~\ref{fig:pv}). 
				CALYPSO maps in this velocity range {(Fig.~\ref{fig:EHV-IHV-SHV})} reveal
				U-shaped cavities opening away from the source in CO, 
				and broad extended wings {behind} arcs (\textit{b}) and (\textit{c}) in SiO and SO, 
				suggestive of material entrained by jet bowshocks.} 				
			
				\vspace{-0.15cm}	
			\subsection{Jet wiggling {and limits on jet rotation}}
				\vspace{-0.03cm}	
			
			Figure~\ref{fig:siocentroid} shows a slightly enlarged region of the blueshifted jet in CO and SiO. 
			It reveals that \tcr{the S-shaped morphology seen in H$_2$ (HC14) is part of} a \tcr{coherent} wiggling pattern traced in {SiO and} CO out to 5-6\arcsec, with a (projected) half-amplitude $\alpha_{\rm obs}$\,$\simeq$\,10\degr\,and wavelength $\lambda_{\rm obs}$\,$\simeq$\,\tcr{3\arcsec\ (700\,au), shorter than the 8.5\arcsec suggested by \citet{2016Chen}.}
		\tcr{We  show in Fig.~\ref{fig:siocentroid} that {this pattern is very well reproduced by} simple models of circular orbital motion 
		or jet precession {of period $\simeq$\,100\,yr} (see Appendix~\ref{app:wiggling} for {details})}.}  
		\tcr{In the orbital scenario, a separation of 70 au (i.e. VLA4A) {would imply, through Kepler's third law,} a total mass $M_{\rm tot}$\,$\simeq$\,7-35\,$M_\odot$ (for eccentricity $e$=0.8\,-\,0),
		{which} is excluded. {A} realistic $M_{\rm tot}$\,$<$\,4\,$M_\odot$ {would thus require} a (so far unseen) closer companion 20--{32}\,au from VLA4B, making SVS13A a hierarchical triple. Its non-detection in NIR images (HC14) implies it would be substantially fainter than VLA4B.} 
			\tcr{Alternatively, CO jet wiggling could be due to jet precession, {with the physical cause remaining unclear:} 
			{rigid disk precession induced by a planetary mass body at 0.5\,au, or non-rigid precession of a larger disk, 
			for which models are still lacking (see Appendix C).}
{Both the orbiting and precessing} jet models originating from VLA4B {reproduce} the [FeII] micro-jet observed by HC14 within 0.3\arcsec,  
{while if the CO jet originates from a companion to VLA4B,} the truncation of the [FeII] jet might be {explained by} jet collision near the H$_2$ bubble center
{(see zoom in Fig.~\ref{fig:Hodapp_zoom})}.}
			
			\FigSiOcentroid 
			
						We stress that the combination of CO jet wiggling and velocity variability will create transverse velocity gradients, that could be mistaken for jet rotation \tcr{at low angular resolution}.  A likely example is the gradient reported by \citet{2016Chen} across ``Bullet~2", the EHV CO bullet located $\simeq$10\arcsec\ further south (see Fig.~\ref{fig:12CO_widefield}): \tcr{its lateral extent ($\pm$\,2.5\arcsec) matches {precisely} the extrapolation of small-scale wiggling ($\alpha_{\rm obs}$\,$\simeq$\,10\degr\,) and its velocity range ({$V_{LSR} \simeq$} -105 to -90\,\kms)} is typical of radial velocity fluctuations {along} the inner EHV jet (Sect. \ref{sect:ori_kine}).
			To constrain the level of rotation in the EHV jet {with minimal confusion from wiggling}, we present in Fig.~\ref{fig:siocentroid} (right panels) transverse PV cuts in CO and SiO within 1.8$\arcsec$ of the source.  {No consistent transverse gradient is found at our resolution. The  jet radius $\le$\,0.4\arcsec and 
				velocity width $<$\,10\,\kms, corrected for an inclination of 20\degr, set a conservative upper limit to the specific angular momentum
				of 1500\,au\,\kms, 4 times smaller than \citet{2016Chen}. 
				If the jet is {steady, axisymmetric, and} magneto-centrifugally accelerated to 100\,\kms\ from a  {Keplerian disk}, the launch radius would be {$<$\,4\,au\,$\times$\,$(M_\star/M_\odot)^{1/3}$ \citep{2003Anderson}}. 	
					
						\vspace{-0.13cm}
				\subsection{Evidence for multiple jets and origin of HH7--11}
						\vspace{-0.03cm}				
				
				\Figurefour 
				
				\tcr{B00 noted that the CO EHV jet at PA\,$\simeq$\,155\degr\, lies in a markedly different direction from the chain of optical Herbig-Haro (hereafter HH) objects HH7-11, at PA\,$\simeq$\,130\degr,} \tcr{and interpreted it as a recent change of ejection direction in a single jet.} 
				Below we argue that these two flows are actually driven by two distinct sources.
				\tcr{First, ejection along HH7-11 still seems active:}		
				Figure~\ref{fig:siocentroid}\tcr{(c)} reveals an unresolved SiO knot of \tcr{unknown origin}
				\tcr{($\ge$\,15$\sigma$ in both SiO and SO, and denoted "knot X" hereafter)} at \tcr{a smaller PA}
				and much lower radial velocity (\tcr{-33 \kms$<$\,$V_{LSR}$\,$<$\,-20\,\kms)} than the local EHV CO jet.
				{These properties} are more typical of H$_2$ emission towards {the wings of} {HH10--11}  \citep{2000Davis, 2001Davis}, 
				suggesting a new event in the HH chain. \tcr{A faint hint of EHV CO towards HH11 is also seen closer in
				Fig. \ref{fig:12CO_widefield}.}			
				\tcr{Second, we find that EHV CO ``bullets'' are not younger but coeval with HH7--11 knots.} We note a close pairing in distance
				between HH11 and CO Bullet~2 (see Fig.~\ref{fig:12CO_widefield}), and between HH10 and the third CO Bullet \tcr{at $\simeq$\,25\arcsec} (Fig.~1 of B00).
				Likewise,  the knot X lies right next to a striking local broadening of the EHV CO jet, that {may} signal a nascent CO bullet.
				{Since the two jets have similar \tcr{projected extents, and also similar} proper motions:} 26-52\,\kms\ for HH10-HH11
				\citep[][scaled to 235~pc]{2001Noriega}, 35\,\kms\ for the H$_2$ arc (HC14) {we {are forced to} conclude that
				HH7-11 knots and CO bullets trace quasi-simultaneous \tcr{recurrent} events occurring in two different directions \tcr{(in projection)}, i.e. two distinct jets from two distinct sources. \tcr{Their jet/disk orientations differ only by 9\degr\ in 3D}\footnote{\tcr{$i$\,$\simeq$\,20\degr\ for the CO jet, $i$\,$\simeq$\,16\degr\ for HH11 {\citep[][scaled to 235~pc]{2001Noriega}} and $\Delta {\rm PA}$\,$\simeq$\, 20\degr\ {yield $\Delta\theta$(3D)\,=\,9\degr}}}.
				
				\tcr{We also report tentative} evidence for a third jet from SVS\,13A.
					\tcr{As shown by the dark grey curve in Figure~\ref{fig:12CO_widefield}}, a curved {chain} of red-shifted knots is present in the blue outflow lobe, and a symmetric blue-shifted chain in the northern red lobe.					
					Such velocity sign reversal {is not expected for outflow} rotation about the VLA4B jet axis\footnote{it {requires} a ratio of rotation to poloidal speeds $V_\phi / V_p  > 1/\tan(i) \simeq 3$, unlike current MHD wind models \citep{2000Casse}.}. Hence, it might trace a third jet {at PA\,$\simeq$\,0\degr}. 
					
							\vspace{-0.4cm}
					\section{\tcr{Discussion and} conclusions}\label{sect:conclusions}

\noindent The CALYPSO view reveals the coexistence of at least two, and possibly three distinct jets from SVS13\,A =VLA4A/4B : (1) a wiggling CO/H$_2$ jet of mean PA\,$\simeq$\,155$\degr$, confirmed to originate from the vicinity of VLA4B {(as suggested by HC14)}; (2) the HH7--11 chain, quasi-synchronous with CO jet bullets but {in a} different PA $\simeq$\,130-140$\degr$; (3) a tentative jet {at PA\,$\simeq$\,0$\degr$}. 
In line with this, SVS13A is found to host two, and up to three possible jet sources: VLA4A (where we find $0.01M_\odot$ of dusty material within 50\,au, and hot corino emission), the NIR source VLA4B, and a {putative} 20--30 au stellar companion \tcr{VLA4Bc, if CO/SiO jet wiggling is due to orbital motion.}

\tcr{The close pairing we uncover between major ejection episodes in the \tcr{"twin"} CO and HH7-11 jets implies a {common} external trigger, making SVS13\,A a strong case for the role of stellar encounters in accretion/ejection bursts in young stars \citep{2014Reipurth,2014Audard}.} Based on the event period $P$\,$\simeq$\,300\,yr \tcr{(bullet projected separation $\simeq$\,10\arcsec), we propose that {synchronous} disk instabilities are} triggered by close perihelion approach of VLA4A in an eccentric orbit around VLA4B. Kepler's third law yields an orbit semi-major axis $a$\,$\simeq$\,45\,$(M_{\rm tot}/M_\odot)^{1/3}(P/300 {\rm yr})^{2/3}$\,au, consistent with this scenario. 
\tcr{No small-scale jet being currently visible from VLA4A, the {origin} of HH7--11 remains ambiguous. It could be episodically launched from VLA4A at periastron passage, and the CO jet from VLA4B; or the HH7--11 and CO jets could {both} arise from the {putative} 20-30~au close binary VLA4B/4Bc, with VLA4A acting merely as external destabilizing agent, and possibly driving the third jet at PA\,$\simeq$\,0$\degr$. \tcr{Searches for the tentative companion are needed to pin-point the source of each jet in this complex system.} \tcr{In either scenario, SVS\,13A is launching {the twin CO and HH7-11} jets from quasi-coplanar disks\footnotemark[3] separated by 20-70\,au, making it highly reminiscent of L1551-IRS5 \citep{2006Lim-Takakuwa}.} }

				\Online
				\begin{appendix}	
					
					\section{Revised proper motions}\label{app:pmotion}

\begin{table}[b!]
	\caption{Previous and revised proper motions}
	\label{tab:pmotion}
	\begin{tabular}{|c|c|c|c|c|}
		\hline
		& \multicolumn{2}{|c|}{CG08\tablefootmark{1}} & \multicolumn{2}{|c|}{This work}\\
		\hline
		Source & $\mu_\alpha$cos$\delta$ & $\mu_\delta$ & $\mu_\alpha$cos$\delta$ & $\mu_\delta$ \\
		& mas yr$^{-1}$ & mas yr$^{-1}$ & mas yr$^{-1}$ & mas yr$^{-1}$ \\
		\hline
		NGC1333$_\mathrm{VLA}$ & 9 $\pm$ 1 & --9 $\pm$ 1 & & \\ 
		VLA\,4A & 6 $\pm$ 3 & --4 $\pm$ 4 & 8.6 $\pm$ 1 & --9.1 $\pm$ 1\\
		VLA\,4B & 7 $\pm$ 3 & --13 $\pm$ 4 & 7.4 $\pm$ 1 & --10.0 $\pm$ 1 \\
		VLA\,4A+B & 6.5 $\pm$ 3 & --8.5 $\pm$ 4  & 8.0 $\pm$ 0.7 & --9.6 $\pm$ 0.7\\
		\hline
		\hline
	\end{tabular}
	\\
	Refs:
	\tablefoottext{1}{\cite{2008Carrasco-Gonzalez}}
\end{table}

					\Figpmotion
					
					Published proper motions for \tcr{VLA4A and VLA4B} were obtained by \citet[][hereafter CG08]{2008Carrasco-Gonzalez} \tcr{from 3.6\,cm observations} \tcr{covering the period 1989-2000}. 
					\tcr{Fig. \ref{fig:pmotion} shows that they} \tcr{do not extrapolate exactly to more recent 8--10\,mm positions of  \citet{2016Tobin},} \tcr{which have a stronger contribution from circumstellar dust making them more relevant for comparison with 1.4\,mm CALYPSO data.} 
					\tcr{Fig. \ref{fig:pmotion} shows no evidence for differing proper motions between 3.6cm (pure free-free emission) and $\lambda \le$\,10\,mm where dust contributes.} Indeed, SVS13A is not affected by contamination from free-free emission because the position of the 3.6\,cm and 7\,mm observations are in agreement within 20\,mas (as shown by \citealt{2004Anglada}), which is also fully compatible with the PdBI astrometric precision. We therefore computed revised proper motions \tcr{by a linear fit to all VLA radio data} \tcr{(see Fig. \ref{fig:pmotion}).}  \tcr{The new values agree better with the average proper motion of VLA sources in the NGC1333 region} \tcr{(see Table \ref{tab:pmotion})}. We deduced the positions at CALYPSO epoch (2012.43) to be: (03:29:03.7416, +31:16:03.812) for VLA4A and (03:29:03.7648, +31:16:03.834) for VLA4B (J2000). \tcr{Using only data at $\lambda$\,$\le$\,10\,mm for proper motions would not significantly change our results.} 
					\vspace{-0.5cm}

					\section{Source of the NIR continuum and jet}\label{app:jetorigin}
					
					As noted by HC14, the NIR position measured by 2MASS in 1999 better agrees with VLA4B than VLA4A (Fig.~\ref{fig:pmotion}, pink stars), hence VLA4B is the source of the $\simeq$\,1990 outburst. Assuming that the dominant source in the NIR remained the same in 2011-2013 as in 1999, HC14 concluded that VLA4B was also the source of the [FeII] jet and H$_2$  {outflow features} \tcr{(reproduced and sketched in Fig.~\ref{fig:Hodapp_zoom}).} However, given the intrinsic variability of young stars and the fact that VLA4B is in decline phase, one cannot totally exclude VLA4A as the dominant NIR source in 2012. Our CALYPSO data have sufficient angular resolution,  {sensitivity}, and astrometric quality to check this independently. \tcr{This test is shown in Fig.~\ref{fig:Hodapp_EHV} on a single channel map.} The morphology of the CO, SiO, and SO emission \tcr{shows a systematic shift from H$_2$ features when the NIR source is registered on VLA4A (top row of Fig.~\ref{fig:Hodapp_EHV}), whereas they} match very well when the NIR source is registered on VLA4B (bottom row of Fig.~\ref{fig:Hodapp_EHV}). \tcr{This good match is also seen in other} \tcr{individual channel maps in Appendix \ref{app:cmaps}, and in the integrated velocity ranges in Fig.~\ref{fig:EHV-IHV-SHV}. Implications are discussed in Section \ref{sect:ori_kine}.}

					\Figuretwoapp

					\section{Wiggling models for the CO/H$_2$ jet}
					\label{app:wiggling}
					
					\FiguretwoappHodapp
					
					We computed simple {ballistic} models of CO/H$_2$ jet wiggling in three scenarios : (1) orbital motion {of} VLA4B as the CO jet source, (2) orbital motion {of} a (so far unseen) companion of VLA4B as the CO jet source, (3) precessing jet from VLA4B. We used the equations of \citet{2002Masciadri}, where we added {projection effects at small view angles}. We explored a grid of parameters to find the smallest mean $\chi^2$ deviation from the H$_2$ linear feature, the apex of arc {\textit{(b)}}, and the EHV SiO peak ahead of arc {\textit{(c)}} (see Fig. \ref{fig:Hodapp_zoom}). {The CO EHV jet at 4\arcsec\ to 6\arcsec\ was not used as a constraint in the $\chi^2$, as numerical simulations  show increased deviations from ballistic trajectories beyond the first {wiggle} wavelength \citep{2002Masciadri}, but it was used to constrain the wiggling period.} Given the complex "loop" effects at small view angle, the temporal variation of jet velocity {could not} be inverted separately from the trajectory like for edge--on systems. {Since a} global inversion procedure lies well beyond the scope of this Letter, we approximated the CO jet velocity variations by a simple piece-wise model, with $V_\mathrm{jet}$\,=\,$V_{LSR}$\,--\,$V_\mathrm{sys}$\,=\,-98.5\,\kms\ out to {$\simeq$\,3\arcsec}, and  
$V_{jet}$\,=\,-123.5\,\kms\ from 4 to 6\arcsec, {based on Fig.\ref{fig:pv} and discussion in Section \ref{sect:ori_kine}}. 
To restrict parameter space, {we fixed $i = 20\degr$ and considered} only circular orbits perpendicular to the jet axis (for the orbital case). 
					
We could find a good match to the observed H$_2$/CO jet wiggling in each of the 3 scenarios.
Our best models are plotted in Figure \ref{fig:Hodapp_zoom} on top of the H$_2$ jet features from Fig.~4 of HC14, and in Figure \ref{fig:siocentroid} on top of the CO and SiO EHV integrated maps. The good {match} despite our simplifying assumptions suggests that they should not be too far from the actual solution. The corresponding parameters are listed in Table~\ref{tab:wiggling}. 
 
We checked \textit{a posteriori} if the [Fe~II] micro-jet extending 0.3\arcsec\ from VLA4B at PA\,=\,145\degr\ (HC14, black {segment in}  Fig.~\ref{fig:Hodapp_zoom}) is consistent with our models. We find that our jet models arising from VLA4B automatically reproduce it to within a few degrees (green and orange in Fig.~\ref{fig:Hodapp_zoom}, where we took $V_{jet}$\,=\,-148.5\,\kms\ within 0.3\arcsec, cf. HC14). Alternatively, when the H$_2$/CO jet comes from a companion, our best model {(see red curve in Fig.~\ref{fig:Hodapp_zoom})} intersects in projection the [FeII] jet near the center of inner H$_2$ bubble (a). Both the H$_2$ bubble and the sudden disappearance of the [Fe~II] jet {might then result from} jet collision. {In conclusion,} the [Fe~II] jet coming from VLA4B {appears} consistent with all 3 scenarios.

{Orbital models of CO jet wiggling clearly require a closer (so far unseen) companion to VLA4B than VLA4A.}
A separation of 70\,au (ie. VLA4A) with an orbital period $\simeq$\,{93}\,yr would imply a total mass $M_{\rm tot}$\,$\simeq$\,7-35\,$M_\odot$ (for eccentricity $e = 0.8-0$)
that is clearly excluded. {This directly stems from Kepler's third law $M_{\rm tot}/M_\odot$\,=\,$(a$/au)$^3$\,$(\tau_o$/yr)$^{-2}$ with a semi-major axis $a$\,$\ge$\,$D/(1+e)$}. 
The second part of Table~\ref{tab:wiggling} lists the constraints on binary separation and individual masses 
{for a more realistic $M_{\rm tot}$\,$<$\,4\,$M_\odot$ (compatible with the observed luminosity) and
a minimum CO jet source mass $M_{\rm COjet}$\,$\ge$\,0.1$M_\odot$}. 
A companion separation 20--32~au is needed, and the CO jet source would be the less massive. {The predicted companion positions are marked in Fig.~\ref{fig:Hodapp_zoom}.}

{If CO jet wiggling is instead due to jet precession,}
the current popular interpretation proposed by \citet{1999Terquem} is that it results from inner disk precession induced by a companion on an inclined orbit. 
In the case of rigid precession of a uniform disk, {truncated at $R$\,$\simeq$\,$a/3$}, Equ. (1) of \citet{1999Terquem} 
{relates the precession period  $\tau_p$ to the orbital period as $\tau_o$\,=\,$0.1 \tau_p \mu / \sqrt{1-\mu}$ with
	$\mu$\,=\,$M_2/M_{\rm tot}$.}  
For a realistic CO jet source mass $0.1$\,$\le$\,$M_{\rm COjet}/M_\odot$\,$\le$\,2, with orbital velocity $V_o$\,$\le$\,1.2\kms\ ({orbital wiggles} $\le$\,2\degr) 
we infer a planet-mass companion at $\le$\,0.5\,au {(see Table~\ref{tab:wiggling})}. {However, a disk truncated at 0.15\,au seems unlikely to produce a bright CO jet.} 
{The condition for rigid precession \citep{1999Terquem} translates here as $(R/{\rm au})^{3/2}$\,$<$\,$(H/0.01R)(\tau_p/100{\rm yr})(M_{\rm COjet}/M_\odot)^{1/2}$, with $H$ the scale height. Hence,} \textit{non-rigid} precession of a larger disk induced eg. by VLA4A cannot be excluded, {by lack of predictions in this regime.}

\vspace{-0.5cm}
					
\begin{table}[ht!]
	\caption{Best models for $i=20\degr$ and inferred source properties}
	\label{tab:wiggling}
	\begin{tabular}{|l|c|c|c|}
		\hline
		{Scenario} & 1 & 2 & 3  \\
		{Plotting} color & Orange & Red & Green \\
		Wiggling cause & Orbital & Orbital & Precession  \\
		CO jet source & VLA4B & Companion & VLA4B \\
		
		[FeII] jet source & VLA4B & VLA4B &VLA4B \\
		\hline
		\multicolumn{4}{|c|}{{Wiggling model parameters (bold face are imposed)}}\\ 
		\hline
		{Orbit/precession} sense & \multicolumn{3}{|c|}{Counterclockwise} \\
		Mean jet PA ($\pm$2\degr) & 152\degr & 155\degr& 154\degr \\
		Projected half-angle & 10\degr & 10\degr & 10\degr \\
		\hline
		Orbital period $\tau_o$ (yr) & 92 & 94 & \\
		Orbital velocity $V_o$ (km/s) & 5.8 & 5.8 & $\le$\,\textbf{1.2} \\
		Orbital radius $r_o$ (au) & 18 & 18 &  \\
		PA to companion ($\pm$3\degr) & 249 & 69 & --  \\
		\hline
		Precession period $\tau_p$ (yr) & & & 116  \\
		{Deprojected} half-angle $\delta$ & & & 3.5\degr \\
		\hline
		\multicolumn{4}{|c|}{{Inferred} binary parameters\tablefootmark{a,b} (bold face are imposed)}\\ 
		\hline 
		$a$ (au)                      & 20 -- 32 &  20--\textbf{26}\tablefootmark{c}  & $\le 0.5$ \\
				$M_{\rm tot}$  ($M_\odot$)             & 1--\textbf{4} &  1--2  & 0.1--2  \\
		$M_{\rm COjet}$   ($M_\odot$)    &  \textbf{0.1} -- 1.8  & \textbf{0.1} -- 0.6  &  \textbf{0.1 -- 2} \\
		$M_{\rm 2}$ ($M_\odot$)       & 0.9 -- 2.2 & 0.9 -- 1.4  &  $\le$\,0.01-0.03 \\
		\hline
	\end{tabular}
	\\
	Notes: 
	\tablefoottext{a}{inferred from $M_{\rm tot} = a^3 / \tau_o^2$ and $\mu = M_2 /M_{\rm tot} = r_o / a$}; 
	\tablefoottext{b}{in Scenario 3, {assume  
	rigid disk precession obeying Equ.(1) of \citet{1999Terquem}} with disk radius $R/a \simeq$\,1/3 and $V_o = 2\pi r_o / \tau_o \le 1.2$ \kms\ (jet orbital spreading $< 2\degr$). {However, non-rigid disk precession induced by VLA4A is not excluded.}}
	\tablefoottext{c}{upper value suggested by our model fit.}
\end{table}					
				
\vspace{-1.0cm}

					\section{Acknowledgements}
					
					{We are grateful to C. Dougados for her help with jet wiggling models,} {and to A. Noriega-Crespo for providing his HST image of HH\,7--11.} CALYPSO observations were obtained at the IRAM PdBI (program numbers ul52 and ui52). IRAM is supported by INSU/CNRS (France), MPG (Germany), and IGN (Spain). This project was supported in part by the French programs "Physique et Chimie du Milieu Interstellaire" (PCMI), and "Action Spécifique ALMA" (ASA) funded by the CNRS and CNES, and by the Conseil Scientifique of Observatoire de Paris. It also benefited from support by the European Research Council under the European Union's Seventh Framework Programme (ERC Grant Agreements No. 291294 -'ORISTARS' and No. 679937 -'MagneticYSOs').

					\section{Channel maps of CO, SiO, and SO}\label{app:cmaps}
					\FigHodappCO
					\newpage	
					\FigHodappSiO
					\newpage	
					\FigHodappSO
					
				\end{appendix}		
				
			\end{document}